\documentclass[notitlepage,12pt,groupedaddress,secnumarabic,prd,showkeys,onecolumn,nofootinbib,noshowpacs]{revtex4}
\usepackage{amsfonts}
\usepackage{amsmath}
\usepackage{amssymb}

\begin{document}

\title{Effects of a modified Reissner-Nordstr\"{o}m spacetime}
\author{J.R. Morris}
\affiliation{Physics Department, Indiana University Northwest, 3400 Broadway, Gary, IN
46408, USA}
\email{jmorris@iun.edu}

\begin{abstract}
A modified Reissner-Nordstr\"{o}m spacetime is considered here, where the
central object (for example, a black hole or naked singularity) possesses a
mass, with an \textquotedblleft ordinary\textquotedblright, i.e., Standard
Model (SM) electric charge, along with a \textquotedblleft
dark\textquotedblright\ electric charge associated with dark matter (DM).
The inclusion of this dark charge modifies the gravitational properties of
the spacetime, while not affecting ordinary electrodynamics. Geodesic
motions of both charged and uncharged SM test particles are modified due to
the presence of dark charge. The modifications may allow the detection and
measrement of the dark charge. In particular, (1) the effective potential
for orbiting particles is modified, and consequently, (2) the angular
momenta and ISCOs of test masses in circular orbits are changed from those
for the usual Reissner-Nordstr\"{o}m case. (3) In the case of a naked
singularity, it is possible that a \textquotedblleft levitating
atmosphere\textquotedblright, due to short distance repulsive gravity, forms
at the zero gravity radius, which depends upon both SM and DM charges. A
levitating atmosphere may then cloak the naked singularity.
\end{abstract}

\maketitle

\section{Introduction}

\ \ The nature of dark matter remains somewhat mysterious, and its
interaction with ordinary matter, that is, Standard Model (SM) matter, seems
to be extremely weak or nonexistent, except via gravitational interactions.
It is therefore reasonable to entertain the distinct possibility that dark
matter (DM) may reside in the interiors of objects such as black holes or
naked singularities. The types of DM particles and their mutual
interactions, i.e., symmetries, is also unknown. One possibility is that the
DM sector contains an unbroken $U(1)$ abelian gauge symmetry associated with
a \textquotedblleft dark charge\textquotedblright\ and a massless
\textquotedblleft dark photon\textquotedblright, which may be completely
decoupled from the SM sector.

\bigskip

\ \ It therefore seems worthwhile to examine the possibility of a modified
Reissner-Nordstr\"{o}m (RN) spacetime. Specifically, we assume the
coexistence of Standard Model (SM), or `ordinary' matter along with `dark'
matter (DM) occupying the same region of space with metric $%
g_{\mu\nu}(x^{\alpha})$. The source (assumed to be some compact object such
as a black hole or naked singularity) has a total mass $M$ and a total
\textquotedblleft electric\textquotedblright\ charge $Q$ composed of an
unspecified proportion of SM and DM matter. The total mass is $M=M_{S}+M_{D}$
and the total charge is $Q_{Tot}=Q_{S}+Q_{D}$, where $Q_{S(D)}$ is the SM
(DM) $U(1)$ charge associated with an unbroken $U(1)_{S}\times U(1)_{D}$
group (i.e., the associated photons are massless). This modified
Reissner-Nordstr\"{o}m metric depends only on the total mass $M$ and a
(charge)$^{2}$ term proportional to $(Q_{S}^{2}+Q_{D}^{2})$, and differs
from the usual case by replacing $Q_{S}^{2}\rightarrow(Q_{S}^{2}+Q_{D}^{2})$
in $g_{\mu\nu}$. It is assumed that the SM and DM sectors are completely
decoupled from each other, i.e., noninteracting, other than gravitationally.
But both SM and DM respond to gravitation in the same way, obeying the
Einstein equations $G_{\mu\nu
}=-\kappa^{2}T_{\mu\nu}^{Tot}=-\kappa^{2}(T_{\mu\nu}^{SM}+T_{\mu\nu}^{DM})$.
(It is also assumed that both SM and DM sectors have the same gravitational
coupling, i.e., $\kappa_{S}=\kappa_{D}=\kappa=\sqrt{8\pi G}$.) However,
although a charge-neutral test mass $m$ responds to the total mass $M$ and
total (charge)$^{2}$ $(Q_{S}^{2}+Q_{D}^{2})$, an ordinary (SM) test charge $%
q $ responds electromagnetically only to the SM charge $Q_{S}$. In this
case, SM charged particle geodesics differ from the normal case where $%
Q_{D}=0$, since the modified RN metric $g_{\mu\nu}(x)$ differs from the
ordinary RN metric for which $Q_{D}=0$. This is due to the presence of the
electromagnetic force $qF_{\mu\nu}^{S}u^{\nu}$ appearing in the geodesic
equations, where $q$ responds only to the charge $Q_{S}$, but a
gravitational effect is due to both charges, $Q_{S}^{2}+Q_{D}^{2}$. We
therefore anticipate an alteration in the expressions for the orbits of
ordinary SM charged (and neutral) particles in comparison to the geodesics
previously studied for the usual Reissner-Nordstr\"{o}m case \cite{Pugliese
PRD11a},\cite{Pugliese PRD11b},\cite{Pugliese EPJC17},\cite{Pugliese MG12}
where $Q=Q_{S}$ and $Q_{D}=0$. In particular, we focus upon SM particles in
circular orbits.

\bigskip

\ \ \ The idea has been entertained that on astrophysical scales, large
masses, and possibly quite large (SM) charges $Q_{S}$, may be encountered
due to different effects. In fact, it has been speculated that enormous
charges on the order of $\sim10^{20}$ C could be supported by highly charged
compact stars, such as neutron stars \cite{Ray PRD03},\cite{Ray BJP04} or
white dwarfs \cite{Carvalho EPJC18}. Even if $Q_{S}$ is small, it has been
speculated \cite{Bai PRD20} that if a sufficiently heavy \textquotedblleft
dark electron\textquotedblright\ exists, the Schwinger effect \cite%
{Schwinger} may be diminished for microscopic primordial black holes,
allowing a large $Q_{D}$.

\bigskip

\ \ A special case is that where the source has a mass $M$, and $Q_{S}=0$
but $Q_{D}\neq0$. In this case the source appears to all ordinary, SM
particles to be a charge-neutral gravitating object with an extra energy
density due to the dark electric field energy density. The modification of
the usual single-charged Reissner-Nordstr\"{o}m metric to the double-charged
metric involves a simple substitution $Q_{S}^{2}\rightarrow
Q_{S}^{2}+Q_{D}^{2}$. Gravitational effects on orbiting objects due to
charge of the source are therefore modified if $Q_{D}\neq0$. This results in
changes in angular momenta $L(r)$, and therefore the radius of an innermost
stable circular orbit (ISCO) for an object orbiting a black hole, and a
change in the \textquotedblleft zero gravity radius\textquotedblright\ of a
spherical surface surrounding a naked singularity (see, for example, \cite%
{Vieira21},\cite{Vieira23}). A zero gravity sphere allows accreted matter to
form a \textquotedblleft levitating atmosphere\textquotedblright, supported
by gravitation and requiring no radiation pressure \cite{Vieira21},\cite%
{Vieira23}. In this sense, the \textquotedblleft levitating
atmosphere\textquotedblright\ around the singularity differs from the
radiation supported \textquotedblleft levitating
atmosphere\textquotedblright\ lying above the surface of a luminous neutron
star \cite{Wielgus1 2015},\cite{Wielgus2 2016}, with both having definite
inner and outer radii.

\section{A modified Reissner-Nordstr\"{o}m metric}

\ \ We assume SM and DM electric charges $Q_{S}$ and $Q_{D}$, respectively,
to be generated by a gauge group $U(1)_{S}\times U(1)_{D}$. An ordinary (SM)
electric charge $q$ interacts only with the $U(1)_{S}$ photon $A_{\mu}$.
(Here, natural units with $\hbar=c=\epsilon_{0}=1$ are used along with
Heaviside-Lorentz electromagnetic units, but $G$ is allowed to remain
visible. We use the $(-,+,-)$ Misner-Thorne-Wheeler classification \cite{MTW}
for the metric, Riemann tensor, and Einstein equation. Conversion to usual
gravitational-Gaussian units amounts to setting $G\rightarrow1$ and
converting the Heaviside-Lorentz charge $Q$ to Gaussian charge $\hat{Q}$
using $\hat
{Q}=Q/\sqrt{4\pi}$.)

\bigskip

\ \ The usual Reissner-Nordstr\"{o}m metric (using the above conventions) is
given by $ds^{2}=f_{0}(r)dt^{2}-f_{0}(r)^{-1}dr^{2}-r^{2}d\Omega^{2}$, ($%
d\Omega^{2}=d\theta^{2}+\sin^{2}\theta d\varphi^{2}$), 
\begin{subequations}
\label{1}
\begin{gather}
f_{0}(r)=g_{00}=-\frac{1}{g_{rr}}=1-\frac{2\mu}{r}+\frac{KQ_{S}^{2}}{r^{2}},
\label{1a} \\
\ \ \mu\equiv GM,\ \ \ KQ_{S}^{2}=\frac{GQ_{S}^{2}}{4\pi},\ \ K\equiv\frac {G%
}{4\pi}   \label{1b}
\end{gather}

where $Q_{S}$ denotes an ordinary (SM) charge, with $KQ_{S}^{2}=\frac {%
GQ_{S}^{2}}{4\pi}$. (For $G\rightarrow1$ and Gaussian units, set $Q=\sqrt{%
4\pi}\hat{Q}$, where $\hat{Q}$ is Gaussian charge, and then $f_{0}(r)=1-%
\frac{2M}{r}+\frac{\hat{Q}_{S}^{2}}{r^{2}}$.)

\bigskip

\ \ The modified RN metric proposed here is of the same form, but with an
additional (dark) $U(1)_{D}$ charge $Q_{D}$ included: $\ f_{0}(r)\rightarrow
f(r)$, with $Q_{S}^{2}\rightarrow(Q_{S}^{2}+Q_{D}^{2})$: 
\end{subequations}
\begin{subequations}
\label{2}
\begin{align}
f(r) & =1-\frac{2\mu}{r}+\frac{Z^{2}}{r^{2}}=1-\frac{2\mu}{r}+\frac {%
K(Q_{S}^{2}+Q_{D}^{2})}{r^{2}},  \label{2a} \\
\ \mu & \equiv GM;\ \ \ \ Z^{2}\equiv K(Q_{S}^{2}+Q_{D}^{2})=\frac{G}{4\pi }%
(Q_{S}^{2}+Q_{D}^{2})   \label{2b}
\end{align}

where $Q_{S}$ and $Q_{D}$ are SM and DM charges, respectively\footnote{$1\ $%
Coulomb $=1.88\times10^{18}$ in natural units ($\hbar=c=1$).}. The $U(1)_{S}$
and $U(1)_{D}$ gauge fields are noninteracting, except through gravitation.
Since the $U(1)$ gauge fields are assumed to be decoupled, the associated
electromagnetic stress-energies are additive, which results in a replacement
of the $Q_{S}^{2}/r^{2}$ term in the usual RN metric by a term $%
(Q_{S}^{2}+Q_{D}^{2})/r^{2}$ in the function $f(r)$ of the modified metric.
Specifically, for the modified RN metric we have $f_{0}(r)\rightarrow f(r)$: 
\end{subequations}
\begin{equation}
\begin{array}{ll}
ds^{2}=f(r)dt^{2}-\dfrac{1}{f(r)}dr^{2}-r^{2}(d\theta^{2}+\sin^{2}\theta
d\varphi^{2}),\smallskip &  \\ 
f(r)\smallskip=g_{00}=-\dfrac{1}{g_{rr}}=\Big(1-\dfrac{2\mu}{r}+\dfrac{Z^{2}%
}{r^{2}}\Big) & 
\end{array}
\label{3}
\end{equation}

with $\mu$ and $Z^{2}$ given by (\ref{2b}).

\bigskip

\ \ The action for the system consists of a field theoretic and a particle
contribution for an \textquotedblleft ordinary\textquotedblright\ (SM) test
particle of mass $m$, $S=S_{f}+S_{p}=\int d^{4}x\sqrt{g}\mathcal{L}_{f}+\int
d\tau L_{p}$, where we denote the SM electromagnetic field by $F_{\mu\nu
}=\partial_{\mu}A_{\nu}-\partial_{\nu}A_{\mu}$ and the SM current density by 
$J^{\mu}$. The corresponding DM field is $\mathcal{F}_{\mu\nu}=\partial_{\mu
}\mathcal{A}_{\nu}-\partial_{\nu}\mathcal{A}_{\mu}$, with a dark current
density $\mathcal{J}^{\mu}$. The total charge of the object which appears in
this modified RN metric is $Q_{Tot}=Q_{S}+Q_{D}$. An ordinary (SM) charge $q$
interacts only with the SM fields $A_{\mu}$ and $F_{\mu\nu}$, although both
SM and DM charges contribute to gravitation through $%
(Q_{S}^{2}+Q_{D}^{2})=Q_{Tot}^{2}-2Q_{S}Q_{D}$. The Lagrangians for the
fields and test particles are given by 
\begin{subequations}
\label{4}
\begin{align}
\ \mathcal{L}_{f} & =-\frac{1}{4}F_{\mu\nu}F^{\mu\nu}-\frac{1}{4}\mathcal{F}%
_{\mu\nu}\mathcal{F}^{\mu\nu}-J^{\nu}A_{\nu}-\mathcal{J}^{\nu }\mathcal{A}%
_{\nu}  \label{4a} \\
L_{p} & =\frac{1}{2}g_{\mu\nu}\dot{x}^{\mu}\dot{x}^{\nu}+\beta\frac{q}{m}%
A_{\mu}\dot{x}^{\mu}   \label{4b}
\end{align}

with the overdot representing differentiation with respect to proper time $%
\tau$ for a massive particle (or an affine parameter for a photon). The
equations of motion for the $U(1)$ gauge fields following from $\mathcal{L}%
_{f}$ are $\nabla_{\mu}F^{\mu\nu}=J^{\nu}$, and$\ \nabla_{\mu}\mathcal{F}%
^{\mu\nu}=\mathcal{J}^{\nu}$ along with the associated Bianchi identities.
The region of interest is outside of the source, where $J^{\nu}=\mathcal{J}%
^{\nu }=0$, and we will assume that the charged (SM + DM) source generates a
static, spherically symmetric SM electric field $\mathbf{E}(r)$ with
vanishing magnetic field, $\mathbf{B}=0$, and a DM electric field $%
\boldsymbol{\mathcal{E}}(r)$ (not felt by SM particles), with a vanishing DM
magnetic field $\mathcal{B}=0$.

\bigskip

\ \ The field equations outside of the source (with Bianchi identities) then
reduce to 
\end{subequations}
\begin{equation}
\nabla_{\mu}F^{\mu\nu}=0,\ \ \nabla_{\mu}\tilde{F}^{\mu\nu}=0,\ \ \ \ \ 
\text{and}\ \ \ \ \ \nabla_{\mu}\mathcal{F}^{\mu\nu}=0,\ \ \nabla_{\mu}%
\mathcal{\mathcal{\tilde{F}}}^{\mu\nu}=0   \label{5}
\end{equation}

where the dual tensors are $\tilde{F}_{\mu\nu}=\frac{1}{2}\epsilon_{\mu
\nu\alpha\beta}F^{\alpha\beta}$ and $\mathcal{\tilde{F}}_{\mu\nu}=\frac{1}{2}%
\epsilon_{\mu\nu\alpha\beta}\mathcal{F}^{\alpha\beta}$. In addition to the
mass $M$ of the source, both the SM and DM fields contribute to gravitation
via a stress-energy tensor $T_{\mu\nu}^{(Tot)}=T_{\mu\nu}^{(SM)}+T_{\mu\nu
}^{(DM)}=T_{\mu\nu}+\mathcal{T}_{\mu\nu}$ due to the gauge fields. Assuming
a static, spherical symmetry,\ the electric fields $E_{r}=F^{r0}$ and $%
\mathcal{E}_{r}=\mathcal{F}^{r0}$ follow from (\ref{5}), e,g,, $\partial
_{\mu}[\sqrt{g}F^{\mu\nu}]=0\implies\partial_{r}(r^{2}F^{r0})=0$, ($g=|\det
g_{\mu\nu}|$):%
\begin{equation}
\begin{array}{cc}
E_{r}=F^{r0}=-F_{r0}=\dfrac{Q_{S}}{4\pi r^{2}}\smallskip\medskip &  \\ 
\mathcal{E}_{r}=\mathcal{F}^{r0}=-\mathcal{F}_{r0}=\dfrac{Q_{D}}{4\pi r^{2}}
& 
\end{array}
\label{6}
\end{equation}

where $Q_{Tot}=Q_{S}+Q_{D}$ is the total $U(1)_{S}\times U(1)_{D}$ charge of
the source containing SM + DM matter and charges. The Einstein equation%
\footnote{%
The stress-energy tensor $T_{\mu\nu}^{Tot}=T_{\mu\nu }+\mathcal{T}_{\mu\nu}$
is traceless, $T^{Tot}=T+\mathcal{T}=0$.} $R_{\mu\nu }=-\kappa^{2}T_{\mu\nu}$
is replaced by $R_{\mu\nu}=-\kappa^{2}(T_{\mu\nu }+\mathcal{T}_{\mu\nu})$.
The additivity of the stress-energy tensors for the abelian gauge fields
outside of the mass $M$ results in the RN metric with the replacement $%
Q_{S}^{2}\rightarrow Q_{S}^{2}+Q_{D}^{2}$.

\bigskip

\ \ For this modified Reissner-Nordstr\"{o}m metric with $f(r)=\Big(1-%
\dfrac {2\mu}{r}+\dfrac{Z^{2}}{r^{2}}\Big)$, horizons are located by $f(r)=0$%
, i.e., 
\begin{equation}
r_{\pm}=\mu\pm\sqrt{\mu^{2}-Z^{2}},\ \ \ (\mu^{2}-Z^{2}\geq0)   \label{7}
\end{equation}

\ \ For $|Z|/\mu<1$ we have a black hole, and for $|Z|/\mu=1$ we have an
extremal black hole. However, for $|Z|/\mu>1$ the metric represents a naked
singularity.

\bigskip

\ \ To be seen from the geodesic equations or the effective potential, in
the case of a naked singularity, a \textquotedblleft zero gravity
sphere\textquotedblright\ will exist for neutral test particles \cite%
{Pugliese PRD11a},\cite{Pugliese PRD11b},\cite{Vieira21},\cite{Vieira23} at
a radius $r_{0}=Z^{2}/\mu$. At this radius the spatial acceleration $%
a_{\mu}=u^{\alpha }\nabla_{\alpha}u_{\mu}$ of a neutral test particle placed
at rest will vanish, according to a static observer. At this radius $r_{0}$,
accreting matter can collect to form a spherical \textquotedblleft
cloak\textquotedblright\ around the singularity \cite{Vieira21},\cite%
{Vieira23} as in the case of the usual Reissner-Nordstr\"{o}m metric \cite%
{Pugliese PRD11a},\cite{Pugliese PRD11b},\cite{Vieira21},\cite{Vieira23}.
For $r<r_{0}$ there is a gravitational repulsion, and for $r>r_{0}$ there is
a gravitational attraction. However, in the present case it is possible for
the singularity to have a vanishing \textquotedblleft
ordinary\textquotedblright\ electric charge, $Q_{S}=0$, but a nonvanishing
dark charge, $Q_{D}\neq0$, thereby appearing to be an electrically neutral
object. For $|Z|/\mu>1$ this object may appear as an electrically neutral
cloaked singularity.

\section{Geodesics}

\ \ We consider the motion of an SM test particle of mass $m$ which may, or
may not, have SM charge $q$ ($q=0$ for a neutral test mass) in the presence
of gravitational and electric fields, sourced by some compact object of
total mass $M=M_{SM}+M_{DM}$ and total charge $Q_{Tot}=Q_{S}+Q_{D}$. The
Lagrangian for a test particle of mass $m$ and charge $q$ in the spherically
symmetric spacetime under consideration is taken to be 
\begin{equation}
L_{p}=\frac{1}{2}g_{\mu\nu}\dot{x}^{\mu}\dot{x}^{\nu}+\frac{q}{m}A_{\mu}\dot{%
x}^{\mu}   \label{8}
\end{equation}

where overdot indicates differentiation with respect to an affine parameter,
such as proper time $\tau$. We assume the particle motion to be restricted
to the equatorial plane, with $\theta=\pi/2$, with $\dot{\theta}=\ddot{\theta%
}=0$. The motion of a test particle is then obtained from $L_{p}$ using the
Euler-Lagrange and constraint equations,%
\begin{equation}
\frac{d}{d\tau}(\frac{\partial L_{p}}{\partial\dot{x}^{\mu}})-\frac{\partial
L_{p}}{\partial x^{\mu}}=0;\ \ \ u_{\alpha}u^{\alpha}=\dot{x}_{\alpha}\dot {x%
}^{\alpha}=\beta=\left\{ 
\begin{array}{cc}
1, & m>0 \\ 
0, & m=0%
\end{array}
\right.   \label{9}
\end{equation}

with $u^{\alpha}=\dot{x}^{\alpha}$ and the condition $u_{\alpha}u^{\alpha
}=\beta$ constrains the trajectory to the particle worldline. We will focus
only on the case of a test particle with nonzero mass ($\beta=1$).

\bigskip

\ \ The resulting equation of motion for the test charge can be written in
the equivalent forms 
\begin{subequations}
\label{10}
\begin{gather}
\frac{d}{d\tau}(mg_{\mu\nu}u^{\nu})-\frac{1}{2}m(\partial_{\mu}g_{\alpha%
\beta })u^{\alpha}u^{\beta}=qF_{\mu\nu}u^{\nu}  \label{10a} \\
\text{or}  \notag \\
\frac{du^{\nu}}{d\tau}+\Gamma_{\alpha\beta}^{\nu}u^{\alpha}u^{\beta}=\frac {q%
}{m}F^{\nu\mu}u_{\mu}   \label{10b}
\end{gather}

where the terms on the right hand side are due to the Lorentz force acting
on the particle. For our metric given by (\ref{3}), the Lagrangian $L_{p}$
is cyclic in the variables $t$ and $\varphi$, implying conservation laws for
energy and angular momentum. Writing $p_{\mu}=mu_{\mu}=mg_{\mu\nu}u^{\nu}$
we have for $\mu=0$ and $\mu=\varphi$, 
\end{subequations}
\begin{subequations}
\label{11}
\begin{align}
\frac{\partial L_{p}}{\partial\dot{x}^{0}} & =p_{0}=mu_{0}+qA_{0}=E:\ \ \ \
\ A_{0}(r)=\int F_{r0}dr=\frac{Q_{S}}{4\pi r}  \label{11a} \\
\frac{\partial L_{p}}{\partial\dot{x}^{\varphi}} & =p_{\varphi}=mu_{\varphi
}=mg_{\varphi\varphi}u^{\varphi}=-L:\ \
L=(mr^{2}\sin\theta)u^{\varphi}=mr^{2}u^{\varphi}   \label{11b}
\end{align}

where $E$ and $L$ are constants. As a reminder, the charge $q$ is assumed to
be an ordinary (SM) charge which interacts only with the charge $Q_{S}$
which generates the SM electric field $\mathbf{E}$. The (SM) potential $%
A_{0} $ is determined from the Maxwell field $F_{r0}=\partial_{r}A_{0}$ with 
$A_{0}(r)=\int F_{r0}dr=Q_{S}/(4\pi r)$.

\section{Effective potential}

\ \ \ \ The assumption $u^{\theta}=u_{\theta}=0$, $\dot{u}_{\theta}=0$,
along with the results from (\ref{11}), 
\end{subequations}
\begin{equation}
u_{0}=\frac{1}{m}(E-qA_{0}),\ \ u_{\varphi}=-\frac{L}{m};\ \ \ \ \ \
u_{r}u^{r}+u_{0}u^{0}+u_{\varphi}u^{\varphi}=1;\ \ \ A_{0}=\frac{Q_{S}}{4\pi
r}   \label{12}
\end{equation}

allows the constraint equation $u_{\alpha}u^{\alpha}=1$ (or $p_{\alpha
}p^{\alpha}=m^{2}$, $p_{\alpha}=mu_{\alpha}$) to be written as%
\begin{equation}
g_{rr}(u^{r})^{2}=1-\frac{g^{00}}{m^{2}}(E-qA_{0})^{2}-g^{\varphi\varphi}%
\frac{L^{2}}{m^{2}}   \label{13}
\end{equation}

or,%
\begin{equation}
(u^{r})^{2}=\frac{1}{|g_{rr}|}\left[ \frac{1}{g_{00}m^{2}}(E-qA_{0})^{2}-%
\frac{L^{2}}{|g_{\varphi\varphi}|m^{2}}-1\right]   \label{14}
\end{equation}

Following \cite{Pugliese PRD11a},\cite{Pugliese PRD11b},\cite{Pugliese
EPJC17},\cite{Pugliese MG12},\cite{OR book},\cite{JM GRG11} we define the
dimensionless effective potential $V(r)$ by $(u^{r})^{2}+V^{2}=E^{2}/m^{2}$.
The radial turning points are given by $V^{2}(r)=E^{2}/m^{2}$, i.e., this
defines the radial coordinates $r$ for which $u^{r}=0$. Therefore $%
(u^{r})^{2}+V^{2}=E^{2}/m^{2}$ implies%
\begin{equation}
V^{2}(r)=\frac{E^{2}}{m^{2}}-(u^{r})^{2}   \label{15}
\end{equation}

where $(u^{r})^{2}$ is given by (\ref{14}). For the static, spherically
symmetric spacetime under consideration, with $\theta$ set to $\theta=\pi/2$
for motion in the equatorial plane and $f(r)$ given by (\ref{3}), then $%
\partial_{\mu}g_{\alpha\beta}=0$ for $\mu\neq r$. Using (\ref{14}) and (\ref%
{15}), with $g_{00}|g_{rr}|=1$,

\begin{equation}
V^{2}=\left( \frac{2qA_{0}}{m}\right) \frac{E}{m}-\frac{q^{2}A_{0}^{2}}{m^{2}%
}+\frac{1}{|g_{rr}|}\left( \frac{L^{2}}{|g_{\varphi\varphi}|m^{2}}+1\right) 
\label{16}
\end{equation}

where $|g_{\varphi \varphi }|$ is evaluated at $\theta =\pi /2$.

\subsection{Circular orbits}

\ \ For circular orbits, $u^{r}=0$, $\dot{u}_{r}=0$, and $V^{2}=E^{2}/m^{2}$%
, so that in this case (\ref{16}) gives solutions (with $V_{\pm }=E_{\pm }/m$%
)%
\begin{equation}
\begin{array}{ll}
V_{\pm } & =\dfrac{qA_{0}}{m}\pm \sqrt{g_{00}\Big(\dfrac{L^{2}}{m^{2}r^{2}}+1%
\Big)}\smallskip =\dfrac{qQ_{S}}{4\pi mr}\pm \sqrt{\Big(1-\dfrac{2\mu }{r}+%
\dfrac{Z^{2}}{r^{2}}\Big)\Big(\dfrac{L^{2}}{m^{2}r^{2}}+1\Big)}%
\end{array}
\label{17}
\end{equation}

For the case that $Q_{D}=0$ this reduces to the effective potential given in
Refs.\cite{Pugliese PRD11a},\cite{Pugliese PRD11b} for a charge $q$ in a
circular orbit about a Reissner-Nordstr\"{o}m black hole\footnote{%
We use Heaviside-Lorentz units here, with $A_{0}=Q_{S}/(4\pi r)$.} with
charge $Q=Q_{S}$. The charge product $qQ_{S}$ could be positive or negative, 
$qQ_{S}=\pm|qQ_{S}|$. From this point forward we only consider the potential 
$V_{+}$, which has an asymptotic value $V_{+}(\infty)=1$ (i.e., $E=m$), and
simply denote it by $V$:%
\begin{equation}
V=\dfrac{qQ_{S}}{4\pi mr}+\sqrt{f(r)\left( \dfrac{L^{2}}{m^{2}r^{2}}%
+1\right) },\ \ \ \ \ f(r)=1-\frac{2\mu}{r}+\frac{Z^{2}}{r^{2}}   \label{18}
\end{equation}

\ \ The angular momentum $L$ of a charge $q$ in a circular orbit of radius $%
r $ is a constant on that particular orbit, but $L$ varies with different
orbital radii $r$, and is determined by the conditions%
\begin{equation}
V(r)=\frac{E_{+}}{m},\ \ \ \Big(\frac{\partial V}{\partial r}\Big)_{L}=0 
\label{19}
\end{equation}

For the special case $qQ_{S}=0$, i.e., either $q=0$ (an SM charge neutral
particle) and/or $Q_{S}=0$ (an SM charge neutral source), applying (\ref{19}%
) to (\ref{18}) yields the result%
\begin{equation}
\frac{L^{2}}{m^{2}}=\frac{r^{2}(\mu r-Z^{2})}{r^{2}-3r\mu+2Z^{2}} 
\label{20}
\end{equation}

where $Z^{2}=K(Q_{S}^{2}+Q_{D}^{2})$. When $Q_{S}=0$, this result holds for
any SM particle -- charged or neutral -- in circular orbit of radius $r$.
This agrees with, and generalizes, previous results for the usual
Reissner-Nordstr\"{o}m case \cite{Pugliese PRD11a} where $Q_{D}=0$.

\bigskip

\ \ The radius $r_{ISCO}$ of the innermost stable circular orbit (ISCO) is
determined by the condition \cite{Song EPJC21},\cite{Song note}%
\begin{equation}
\frac{dL^{2}}{dr}\Big|_{r_{ISCO}}=0   \label{21}
\end{equation}

Differentiating (\ref{20}), (\ref{21}) gives the condition (for $r\neq0$)%
\begin{equation}
\mu r^{3}-6\mu^{2}r^{2}+9\mu Z^{2}r-4Z^{4}=0,\ \ \ \ \ (r=r_{ISCO}) 
\label{22}
\end{equation}

This recovers the condition\footnote{%
See Eq.(8) of Ref.\cite{Pugliese PRD11a}.} found in \cite{Pugliese PRD11a}
for the special case $Q_{D}^{2}=0$. In the limiting case of a Schwarzschild
black hole ($Z=0$) we obtain the Schwarzschild value, $r_{ISCO}=6\mu$. Eq.(%
\ref{22}) can be solved for $Z\neq 0$, but the solutions are quite
complicated (see, e.g., Eq.(12) of Ref.\cite{Pugliese PRD11a} for the
special case $Q_{D}=0$), and are not presented here. For $qQ_{S}\neq0$ the
solutions to (\ref{19}) and (\ref{21}) are even more complicated, and are
not considered here.

\bigskip

\ \ We can note the following points.

\bigskip

\ \ (i) Previous detailed analyses for neutral or charged particles in
circular orbits in a Reissner-Nordstr\"{o}m spacetime can be extended and
generalized to the case of a modified Reissner-Nordstr\"{o}m spacetime,
where both SM and DM charges may comprise the source -- e.g., a black hole
or a naked singularity.

\bigskip

\ \ (ii) A source that appears to be SM charge-neutral ($Q_{S}=0$) and
nonrotating, may in fact, contain an apparent mass function $\mathcal{M}%
(r)=M-Z_{D}^{2}/(2r)$ for the metric function $g_{00}=1-\frac{2\mathcal{M}(r)%
}{r}$. The correction term $-Z_{D}^{2}/r=-KQ_{D}^{2}/r$ (mimicing a negative
mass distribution), disappears asymptotically, with $\mathcal{M}(\infty)=M$,
but has an effect on all orbiting SM particles that differs from that
expected for an ordinary Schwarzschild or Reissner-Nordstr\"{o}m source. The
angular momentum of a particle, described by (\ref{20}), and hence the ISCO,
is somewhat sensitive to the value of $Z^{2}$, which generally depends upon
both $Q_{S}$ and $Q_{D}$.

\bigskip

\ \ (iii) The potential exists for a detection and determination of the set
of parameters $\{\mu,Q_{S}^{2},Q_{D}^{2}\}$ characterizing the (spinless)
source. These parameters $\mu$, $Q_{S}^{2}$, and $Q_{D}^{2}$ can, in
principle, be extracted through measurements by a distant observer of the
orbital velocities of test particles in three different circular orbits of
radii $r_{i}$, $i=1,2,3$, i.e., three independent equations containing $%
\{\mu,Q_{S}^{2},Q_{D}^{2}\}$. A distant observer in an asymptotic region of
space at a coordinate distance $R\gg r_{i}$ (and in the equatorial plane)
could measure the rotational velocities $v_{i}^{\widehat{\varphi}}(r_{i})$
(say, by Doppler shifts) of three different test particles in circular
(equatorial) orbits at radii $r_{i}$ from the center of the source. The
physical rotational velocity $v_{i}^{\widehat{\varphi}}(r_{i})$ (velocity in
the $\varphi$ direction) according to a local static observer at a
coordinate distance $r_{i}$ from the center of the source (origin) is%
\begin{equation}
v_{i}^{\widehat{\varphi}}(r_{i})=\frac{\sqrt{g_{\varphi\varphi}(r_{i})}\
d\varphi_{i}}{\sqrt{g_{00}(r_{i})}\ dt_{i}}=\frac{r_{i}}{\sqrt {g_{00}(r_{i})%
}}\ \omega_{i}(r_{i})=\frac{r_{i}}{\sqrt{g_{00}(r_{i})}}\frac{2\pi}{%
T_{i}(r_{i})}   \label{23}
\end{equation}

where $\omega=d\varphi/dt=2\pi/T$ is simply the angular speed of the
orbiting particle as observed by the distant (asymptotic) observer (for
which $g_{00}=1$). If the distant observer can measure $v_{i}^{\widehat{%
\varphi}}(r)$ (e.g., combined gravitational and velocity Doppler shifts) and 
$\omega(r)=2\pi/T(r)$ (e.g., orbital period), then%
\begin{equation}
g_{00}(r_{i})=1-\frac{2\mu}{r_{i}}+\frac{K(Q_{S}^{2}+Q_{D}^{2})}{r_{i}^{2}}=%
\Big[\frac{r_{i}\omega_{i}(r_{i})}{v_{i}^{\widehat{\varphi}}(r_{i})}\Big]%
^{2},\ \ \ (i=1,2,3)   \label{24}
\end{equation}

Solving these three equations simultaneously then yields (in principle) the
set $\{\mu,Q_{S}^{2},Q_{D}^{2}\}$. (Detecting these parameters, in practice,
is not expected to be easily done, requiring skills of experienced
astronomers able to measure combined gravitational and velocity Doppler
shifts, and deal with a realistic set of orbiting objects.) Other methods
for obtaining estimates or bounds on charges, such as lensing effects, or
parameters of noncircular stellar orbits (as for the S2 star around Sgr A*),
and binary systems, may also exist (see, e.g., Ref.\cite{Iorio} and
references therein), but are not considered here.

\section{Zero gravity spheres around naked singularities}

\ \ From the form of $g_{00}(r)=f(r)=1-2\mu/r+Z^{2}/r^{2}$ we find black
hole solutions for $Z^{2}/\mu^{2}\leq1$ with horizons given in (\ref{7}),
where $g_{00}=0$. On the other hand, for $Z^{2}/\mu^{2}>1$ the metric
describes a naked singularity, with $g_{00}$ being nonzero at all spatial
points. However, in this case an interesting situation arises wherein there
are \textquotedblleft zero gravity spheres\textquotedblright\ (see, e.g.,%
\cite{Pugliese PRD11a},\cite{Pugliese PRD11b},\cite{Vieira21},\cite{Vieira23}
and references within.) A \textquotedblleft zero gravity
sphere\textquotedblright\ is a spherical surface where a particle placed at
rest remains at rest in a state of stable equilibrium, according to a
distant static observer. For the metric considered here, a radius $%
r_{0}=Z^{2}/\mu$ locates the radius of a zero gravity sphere (see below) for
a neutral particle ($q=0$) and/or for a neutral singularity ($Q_{S}=0$). At
radii $r<r_{0}$ a neutral particle experiences a gravitational repulsion,
and for radii $r>r_{0}$ gravitation is attractive. By spherical symmetry,
such particles can spherically accrete near the radius $r_{0}$ of the
singularity, oscillate, and through dissipative processes, settle down, more
or less, near the radius $r_{0}$. After time, a spherical cloud, or a
\textquotedblleft levitating atmosphere\textquotedblright, forms near this
zero gravity sphere (see, for example, Refs. \cite{Vieira21} and \cite%
{Vieira23},\ and references within). A levitating atmosphere around a naked
singularity is the analogue of the levitating atmosphere above the surface
of a luminous neutron star, which is supported by radiation \cite{Wielgus1
2015},\cite{Wielgus2 2016}. However, the levitating atmosphere around the
singularity is supported solely by gravitation, requiring no radiative
support \cite{Vieira21},\cite{Vieira23}. The radius $r_{0}$ of the zero
gravity sphere locates the minimum of the effective potential $V(r)$,
determined by the conditions $\partial _{r}V|_{r_{0}}=0$ and $%
\partial_{r}^{2}V|_{r_{0}}>0$ at the equilibrium point $r_{0}$. \ For $%
Z^{2}/\mu^{2}>1$ this minimum is radially stable.

\bigskip

\ \ Similar statements can be made for charged particles ($q\neq 0$) and a
charged singularity ($Q_{S}\neq 0$), but the solutions for the corresponding
radii $r_{q}^{\pm }$ are a little more complicated. Nevertheless, small
oscillations about these radial positions are expected to become damped, and
charged particles may combine to form neutral ones, allowing the formation
of a levitating cloud, or \textquotedblleft atmosphere\textquotedblright\ to
form around the singularity. The objective here is to simply show the
existence of these zero gravity spheres for the extended Reissner-Nordstr%
\"{o}m metric, where $Z^{2}$ depends upon both $Q_{S}^{2}$ and $Q_{D}^{2}$.
What would normally be considered to be an \textquotedblleft
undercharged\textquotedblright\ Reissner-Nordstr\"{o}m black hole (with $%
Q_{S}^{2}<\mu ^{2}$) may actually be an \textquotedblleft
overcharged\textquotedblright\ extended Reissner-Nordstr\"{o}m singularity,
with $Q_{S}^{2}<\mu ^{2}$ but $Z^{2}>\mu ^{2}$. To locate the radii of one
of these equilibrium spheres, we assume a particle to be at rest at a radius 
$r$ with $u^{r}=u^{\theta }=u^{\varphi }=0$. Then, for a static observer,
the condition $\dot{u}^{r}=0$ for a particle locates the radius $r_{0}$ or $%
r_{q}^{\pm }$. Equivalently, for the effective potential we require $L=0$
and $u^{r}=0$ for the particle, and look for the minimum of $V(r)$.

\subsection{$\mathbf{L=0,\ qQ}_{S}\mathbf{=0}$}

\ \ The effective potential for motion at a constant radius, i.e., circular
motion, is given by (\ref{18}), so that for the case for which either or
both the SM charge $q$ of the test mass, or the charge $Q_{S}$ of the
singularity, vanishes, the effective potential for a particle at rest, $L=0$%
, is%
\begin{equation}
V_{0}(r)=\sqrt{g_{00}}=\sqrt{1-\frac{2\mu }{r}+\frac{Z^{2}}{r^{2}}}
\label{25}
\end{equation}

Since $|Z|/\mu >1$ for a singularity, where $|Z|=\sqrt{K(Q_{S}^{2}+Q_{D}^{2})%
}$, then $g_{00}$ and $V_{0}(r)$ are everywhere nonzero. An extremum of $%
V_{0}$ is determined by $\partial _{r}V_{0}|_{r_{0}}=0$ and stability at
this radius $r_{0}$ requires that the extremum is indeed a minimum, $%
\partial _{r}^{2}V_{0}|_{r_{0}}>0$. Using (\ref{25}) we find $\partial
_{r}V_{0}=\frac{1}{\sqrt{g_{00}}}(\frac{\mu }{r^{2}}-\frac{Z^{2}}{r^{3}})$,
so that 
\begin{equation}
r_{0}=\frac{Z^{2}}{\mu }=\frac{G}{4\pi }\frac{(Q_{S}^{2}+Q_{D}^{2})}{\mu }=%
\frac{1}{4\pi }\frac{(Q_{S}^{2}+Q_{D}^{2})}{M},\ \ \ \text{(zero gravity
radius,\ \ }qQ_{S}=0\text{)}  \label{26}
\end{equation}%
and $\partial _{r}^{2}V\Big|_{r_{0}}=\frac{\mu ^{4}}{Z^{6}\sqrt{1-\frac{\mu
^{2}}{Z^{2}}}}>0$, ensuring radial stability for $\mu ^{2}/Z^{2}<1$.

\subsection{$\mathbf{L=0,\ qQ}_{S}\mathbf{\neq 0}$}

\ \ For this case the effective potential is more complicated. For a
stationary particle with $u_{\varphi}=0$ ($L=0$) and $u^{r}=u^{\theta}=0$,%
\begin{equation}
V(r)=\frac{qQ_{S}}{4\pi mr}+\sqrt{g_{00}},\ \ \text{or}\ \ V(r)=\frac{qQ_{S}%
}{4\pi mr}+V_{0}(r)   \label{27}
\end{equation}

where $V_{0}=\sqrt{g_{00}}$. The extrema are located by 
\begin{equation}
\partial_{r}V|_{r_{q}}=0\implies\Big(-\frac{qQ_{S}}{4\pi mr^{2}}+\partial
_{r}V_{0}\Big)_{r_{q}}=0   \label{28}
\end{equation}
\qquad\ 

Now, for convenience, define the $r$ - independent parameter $C\equiv \frac{%
qQ_{S}}{4\pi m}=\frac{\hat{q}\hat{Q}_{S}}{m}$ which depends upon the charge $%
Q_{S}$ of the singularity and the charge-to-mass ratio $q/m$ of the test
particle. Then the potential $V(r)$ can be written as 
\begin{equation}
V(r)=\frac{C}{r}+\sqrt{1-\frac{2\mu}{r}+\frac{Z^{2}}{r^{2}}},\ \ \ \
(C\equiv \frac{qQ_{S}}{4\pi m},\ \ Z^{2}>\mu^{2})   \label{29}
\end{equation}

(Note that in Gaussian units $C=\hat{q}\hat{Q}_{S}/m$.) The parameter $%
C=\pm|C|$ can be positive or negative. The potential for a charged particle, 
$V$, is that for a neutral particle, $V_{0}=\sqrt{g_{00}}$, shifted by the
term $C/r$. The local extrema $r_{q}$ of $V(r)$ are given by Eq.(\ref{28}),
which, for $C^{2}<\mu^{2}$, has solutions\footnote{%
These extrema are stable minima for $\ C^{2}<\mu^{2}<Z^{2}$, satisfying the
conditions $V^{\prime }(r_{q})=0$ and $V^{\prime\prime}(r_{q})>0$.} given by%
\begin{equation}
r_{q}^{\pm}=\frac{\mu(Z^{2}-C^{2})\pm|C|\sqrt{(Z^{2}-\mu^{2})(Z^{2}-C^{2})}}{%
\mu^{2}-C^{2}}   \label{30}
\end{equation}

where $r^{+}$ is the solution for $C>0$ ($qQ_{S}>0$, electric repulsion) and 
$r^{-}$ is the solution for $C<0$ ($qQ_{S}<0$, electric attraction). For the
case that $|C|\ll\mu$, we have, approximately,%
\begin{equation}
r_{q}^{\pm}\approx r_{0}\pm\zeta,\ \ \ \zeta\equiv\frac{|C|\sqrt{%
Z^{2}(Z^{2}-\mu^{2})}}{\mu^{2}}   \label{31}
\end{equation}

For $C=0$ we recover the solution $r_{0}=Z^{2}/\mu $ for a neutral particle $%
q=0$, and/or $Q_{S}=0$ for the singularity. The equilibrium radius $r_{q}^{-}
$ is a little smaller than $r_{0}$ for electrical attraction, while $%
r_{q}^{+}$ is a little further away for electrical repulsion.

\subsection{\textbf{Levitating atmospheres}}

\ \ Although dark matter particles may also accumulate near some equilibrium
sphere, little, if anything, is really known about their interactions among
other particles. However, particles of ordinary matter will interact with
each other, and some particles can eventually collect near the zero gravity
radius $r_{0}$. If a particle is assumed to have some angular momentum $L$,
then $L$ will minimize with $L=0$ at an ISCO of radius \cite{Song note} $%
r_{ISCO}\approx r_{0}$.

\bigskip

\ \ The effective potential $V_{0}(r)=\sqrt{g_{00}(r)}$ for an
\textquotedblleft ordinary\textquotedblright\ neutral particle ($Q_{S}=0$)
diverges at the singularity ($V_{0}(r)\rightarrow+\infty$ as $r\rightarrow0$
for $Z^{2}>\mu^{2}$), so that no such particle of finite energy actually
reaches it. The interactions of ordinary neutral particles near $r_{0}$ can
dissipate energy, so that a particle with initial energy $E/m\gtrsim
V_{0}(r_{0})$ becoming trapped within the potential well will undergo radial
oscillations about $r_{0}$, and form a spherical \textquotedblleft
levitating atmosphere\textquotedblright\ near that radius. The
characteristics of such atmospheres around naked singularities possessing
zero gravity spheres has been studied previously by Vieira and Klu\'{z}niak
(see, for example, \cite{Vieira21},\cite{Vieira23} and references therein).
In this way the naked singularity becomes \textquotedblleft
cloaked\textquotedblright\ by the levitating atmosphere. It is proposed \cite%
{Vieira23} that such an atmosphere can be formed in a very short time by the
accretion of ambient matter. This \textquotedblleft levitating
atmosphere\textquotedblright\ around the singularity, is analogous to the
radiation supported \textquotedblleft levitating
atmosphere\textquotedblright\ lying above the surface of a luminous neutron
star \cite{Wielgus1 2015},\cite{Wielgus2 2016}, both having definite inner
and outer radii. However, the levitating atmosphere around the singularity
is supported only by gravity, requiring no radiative support.

\bigskip

\ \ The situation for charged particles, however, is less clear, since $%
V(r)=C/r+V_{0}(r)\rightarrow\pm\infty$, depending on the sign and magnitude
of $C\propto qQ_{S}/m$ and $V_{0}(r)$. In the case that $C<0$ (electric
attraction) with $|C|>|Z|$, then $V(r)$ decreases toward the origin and
becomes negative with $V(r)\rightarrow-\infty$ as $r\rightarrow0$. An
unperturbed electric field $E_{r}\sim Q_{S}/r^{2}$ then grows in magnitude
without bound at small $r$. If at some small $r_{c}$ this field reaches a
critical strength $E(r_{c})=E_{c}\sim m_{e}^{2}/e$ where the Schwinger
effect \cite{Schwinger} becomes important, then a cascade of $e^{+}e^{-}$
pairs may be coaxed out of a destabilized vacuum at radii $r\lesssim r_{c}$,
with the ensuing evolution becoming problematic. If enough charged particles
of sign opposite to that of $Q_{S}$ fall onto the singularity, the result
would be a reduction of $|Q_{S}|$ and $|Z|$ and an increase in mass $\mu$,
possibly resulting in a state where $Z^{2}<\mu^{2}$. One then expects the
formation of horizons, resulting in the naked singularity evolving into a
black hole. This process would support the claims of Cohen and Gautreau \cite%
{Cohen PRD79} that a Reissner-Nordstr\"{o}m naked singularity can be
destroyed but not created, while event horizons can be created but not
destroyed.

\section{Conclusions}

\ \ We have considered the simple, but interesting, possibility that some
compact objects -- black holes or naked singularities -- can harbor an
ordinary, Standard Model, charge $Q_{S}$ and/or a dark charge $Q_{D}$ due to
an unbroken $U(1)_{S}\times U(1)_{D}$ symmetry of the source. The
corresponding photons are massless, and the Reissner-Nordstr\"{o}m solution
to the Einstein equation is extended so that the metric components $%
g_{00}(r)=-g^{rr}(r)=f(r)$ depend upon both charges $Q_{S}$ and $Q_{D}$ as
in Eqs.(\ref{2}) and (\ref{3}). The parameter $Z^{2}%
\propto(Q_{S}^{2}+Q_{D}^{2})$ now appears in the metric component $%
g_{00}(r)=f(r)$. For $\mu^{2}>Z^{2}$ the object is a black hole, for $%
\mu^{2}=Z^{2}$ it is an extremal black hole, and for $\mu^{2}<Z^{2}$ the
object is a naked singularity. The dark matter and dark charge are assumed
to be completely decoupled from the SM sector, except via gravitation. The
presence of a nonzero dark charge will alter the geodesics expected for an
ordinary Reissner-Nordstr\"{o}m source. Geodesic motions of both neutral and
charged particles in circular orbits around ordinary Reissner-Nordstr\"{o}m
black holes and naked singularities have been well studied (see, for
example, \cite{Pugliese PRD11a} and \cite{Pugliese PRD11b} and references
therein), but the inclusion of a nonzero dark charge will modify these
previous results with a replacement of $Q_{S}^{2}%
\rightarrow(Q_{S}^{2}+Q_{D}^{2})$ in the metric. In principle, these
modifications could allow the inference of the existence and inclusion of
dark matter and dark charge hidden in the gravitational source.

\bigskip

\ \ For ordinary SM particles in circular orbits around a source with $%
Q_{D}\neq0$, the angular momentum per unit mass $L(r)/m$ differs from that
due to a standard Reissner-Nordstr\"{o}m source. In fact, even in the
extreme case $Q_{S}=0$, all SM particles, charged or neutral, have angular
momenta $L(r)/m$ determined only by the parameters $M$ and $Q_{D}$ of the
source (see (\ref{20})). In addition, the radius $r_{ISCO}$ of the ISCO will
depend upon both types of charge (see (\ref{21}) and (\ref{22})). This
observation for an idealized situation concerning test particles may lead to
insights for a more realistic case where a real accretion disc, with its
complex internal interactions, may have properties dependent upon the dark
charge $Q_{D}$ of the source. Since the metric components $g_{00}$ and $%
g_{rr}$ for the modified metric generally differ from those for the usual
one, the presence of nonzero $Q_{D}$ changes the position and structure of a
black hole horizon or naked singularity.

\bigskip

\ \ In the case of a naked singularity, a \textquotedblleft zero
gravity\textquotedblright\ spherical surface exists, where a particle having
zero angular momentum $L=0$, \cite{Pugliese PRD11a}, \cite{Pugliese PRD11b}
can remain at rest. Consequently, it is possible for a \textquotedblleft
levitating atmosphere\textquotedblright\ of accreted matter\ of the type
previously described by Vieira and Klu\'{z}niak \cite{Vieira21},\cite%
{Vieira23} to form, so that the cloaked singularity may take the appearance
of a more ordinary astrophysical object. This levitating atmosphere
surrounding a naked singularity is analogous to the levitating atmosphere
above a luminous neutron star \cite{Wielgus1 2015},\cite{Wielgus2 2016},
both having a finite thickness. The radius of a zero gravity sphere for a
neutral particle is given by $r_{0}=Z^{2}/\mu$, which generalizes the
results found \cite{Pugliese PRD11a},\cite{Vieira21},\cite{Vieira23} for the
usual Reissner-Nordstr\"{o}m spacetime with $Q_{S}\neq0$, $Q_{D}=0$. Since
the radius of a zero gravity sphere depends upon both charges $Q_{S}$ and $%
Q_{D}$, a possible consequence is that an inferred observational value for $%
Q_{S}$ may actually be an inferred value for $|Z|=\sqrt{%
K(Q_{S}^{2}+Q_{D}^{2})}=\sqrt{\hat{Q}_{S}^{2}+\hat{Q}_{D}^{2}}$.

\bigskip

\ \ The radius of a zero gravity sphere locates the stable minimum of the
effective potential where $u^{r}=u^{\theta}=u^{\varphi}=0$, and a neutral
particle placed at rest there remains at rest. For $r<r_{0}$ there is a
gravitational repulsion, which (for a naked singularity with $Z^{2}>\mu^{2}$%
) diverges to $+\infty$ as $r\rightarrow0$, preventing any massive neutral
particle with finite energy from reaching the singularity. As emphasized by
Mishra and Vieira\cite{Mishra23}, no stable circular orbits with $r<r_{0}$
exist for a naked singularity. If a gravitating source is assumed to be a
naked singularity, an observed lower bound for stable circular orbits can
yield a bound on $r_{0}$ and $|Z|/\mu$. As an interesting example, Mishra
and Vieira \cite{Mishra23} have considered the compact source Sgr A$^{\ast}$
at the center of the galaxy to possibly be a Reissner-Nordstr\"{o}m
singularity. In this case it is argued that, based upon the Event Horizon
Telescope observations \cite{EHTa}, the apparent size of Sgr A$^{\ast}$
allows an inference of its charge to mass ratio to lie within a restricted
range of $1<|\hat{Q}_{S}|/\mu<2.32$ (using geometrized Gaussian units).
However, the replacement $Q_{S}^{2}\rightarrow Z^{2}\rightarrow(\hat{Q}%
_{S}^{2}+\hat{Q}_{D}^{2})$ could be used to obtain bounds for a dark charge $%
1<|\hat{Q}_{D}|/\mu<2.32$ for Sgr A$^{\ast}$ if it is assumed to have $\hat{Q%
}_{S}=0$.

\bigskip

\ \ In conclusion, we have studied the possibility that a gravitating
source, such as a black hole or naked singularity, can possess an abelian $%
U(1)$ symmetry associated with a dark charge, analogous to the corresponding
symmetry of the Standard Model, and the possible consequences. If an abelian
dark charge with an attendant electrodynamics does in fact exist, a new
avenue of probing dark matter properties might be opened. Since the nature
of dark matter remains so enigmatic, theoretical explorations of its
possible properties and effects are essential for further understanding.


\begin{thebibliography}{99}
\bibitem{Song note} Song \cite{Song EPJC21} shows that the effective
potential $V(r)$ for a particle in circular motion will minimize when the
angular momentum minimizes, with an ISCO determined by $%
dL^{2}/dr|_{r_{ISCO}}=0$.
\end{thebibliography}
\end{document}